# Study of Solar Energetic Particles: their Source Regions, Flares and CMEs during Solar Cycles 23-24

*Raj Kumar[1], Ramesh Chandra[1], Bimal Pande[1], Seema Pande[1]

[1]Department of Physics, DSB Campus, Kumaun University, Nainital, India

## ABSTRACT

In this work, we examine the association between solar active regions and 152 solar flares, coronal mass ejections, and solar energetic particle (SEP) events over solar cycles 23–24 (1997–2017). The Coordinated Data Analysis Workshops (CDAW) center's GOES data in the energy channel > 10 MeV (Major SEPs; solar proton events) with flux ≥ 10 pfu was used for our investigation. For the associated activities, we have analyzed the data from space born satellites namely: SOHO/LASCO and SDO/AIA. We found a moderate correlation (55%) between SXR flux and sunspot area i.e., active regions with larger sunspot areas generally generate larger flares. We found that most of the SEPs are originated from the magnetically complex active regions i.e., hale class βγδ and β. Very few events were associated with unipolar active regions. Stronger GOES X-ray is linked to more impulsive events, as evidenced by the negative correlation (-0.40) between X-ray flux and SEP duration. In the active region βγδ, the highest average SEP intensity (2051 pfu) was detected. In the data set used, only 10% SEPs are found impulsive in nature, while the remaining 90% are gradual in nature. All the impulsive events had SEP intensity less than 100 pfu and most of the CMEs associated with these events were decelerated CMEs. We discovered that the majority of faster CMEs are linked to the most complex magnetic active regions. This indicates that high speed CMEs are produced by magnetically complex active regions. We discovered that 58 SEP events in our data set are linked to accelerated CMEs, while 82 are linked to decelerated CMEs. The highest average CME width is found corresponding to magnetically most complex active regions βδ, γδ, αγδ and βγδ, which shows that large CMEs are the consequences of magnetically complex active regions.

***Keywords:*** Solar energetic particle events — Solar active region— Coronal mass ejections — Gradual SEP events — Impulsive SEP events.

*Corresponding Author
Email: rajkchanyal@gmail.com

## 1. Introduction

High-energy particles with energies ranging from a few thousand electron volts to giga electron volts are known as Solar Energetic Particle (SEP) events, whereas Solar Proton Events (SPEs) have energies ranging from a few million electron volts to giga electron volts. Forbush saw the first SEP event in 1946 [1], and it was eventually identified as ground level enhancements [2]. Much earlier than the earliest observable evidence of SEP events Carrington (1859) [3] was the first to report solar flares. Thus, until the 1970s, the SEP events were thought to be a result of solar flares. After the discovery of Coronal Mass Ejections (CMEs) in the 1970s, a different theory for how SEP events are produced emerged [4] i.e., CME shocks generation and they were categorized as either impulsive or gradual SEP events [5–13].

Generally SEP events of very short duration are impulsive SEP events while the SEPs of long duration are termed as gradual SEP events. Gradual events are long duration events ranging from 1 to 3 days, while impulsive events are short lived in nature with the SEP duration of less than one day (< 1-20 h) [14]. Every year, approximately one thousand impulsive events are observed while a few tens of large events are observed per year. In impulsive SEP events, the associated X-ray flares are also impulsive in nature, typically lasting up to an hour, whereas in gradual SEP events, the flare duration exceeds one hour [15, 16]. These two types of events can be categorized on the basis of energetic particle composition and radio observations [11]. Impulsive events are $^{3}$He rich events which are accompanied by type-III radio bursts [17]. These electron-rich events exhibit $^{3}$He/$^{4}$He ratios, ranging from 1,000 to 10,000 and Fe/O ratios up to a factor of 10.

The gradual events are proton rich events which have Fe/O ratio up to 0.1. These events do not have a measurable amount of $^{3}$He/$^{4}$He ratios and are associated with type II radio bursts [13, 18]. The gradual events are observed from both the east and west longitudinal distribution of flares, while the impulsive events are mostly observed from western longitudinal flare sites [13, 14].

According to Gopalswamy *et al.,* 2007 [19], SEPs can be the consequences of energetic CMEs. Generally, gradual SEP events are considered to be originated due to CME driven shocks while impulsive SEP events are originated due to the solar flares associated with magnetic reconnection [13, 14, 20- 23].

Strong SEPs are associated with the stronger shocks which are originated from high speed CMEs [24-26] with average speed of ≈ 1500 km $s^{-1}$ [27] and ≈ 1240 km $s^{-1}$ [28]. Many authors looked for a desired strong correlation between the parameters of CME on one hand and SEPs on another, but didn't find a strong correlation between these parameters [29-33]. According to Gopalswamy *et al.,* 2009 [34], CME–CME interaction, CMEs deflection from the coronal hole regions etc. might be the reasons for the occurrence of this weak correlation.

Another possible mechanism of the impulsive SEPs is flare reconnection process [35, 36]. Numerous statistical studies have shown that the correlation between SEPs parameters and solar X-ray flux is found weak [37, 38]. In a recent study, Kumar, *et al.,* 2020 [28] found a weak correlation (r = 0.40) between SEP intensity and X-ray flux. So it is considered that neither the CME shocks nor the flare reconnection processes are separately responsible for the generation of SEP events, but both the mechanisms as a whole contribute for the same, although their relative contribution is not exactly clear [39, 40].

Magnetic topology of active regions (ARs) with the shock and reconnection mechanisms should also be taken into account for solving the puzzle of SEPs generation mechanism [28, 41-45]. Numerous studies have explored the relationship between active region complexity and the occurrence of solar flares and CMEs, revealing that greater magnetic field complexity increases the likelihood of generating intense flares and large CMEs [46, 47]. The large active regions having more complex magnetic structures and sufficient stored magnetic energy are responsible for the generation of strong flares and CMEs [42, 48-50].

We therefore present here a study of SEPs, impulsive and gradual SEP events with their linked source regions, associated CMEs and soft X-ray flares. The relative contribution of CMEs, solar flares and magnetic complexity of source regions to the occurrence of SEPs is described with their correlation study. Distribution of SEP intensity, X-ray flux, CME speed, CME acceleration and deceleration is examined with different type of magnetically complex (Hale classes) active regions. Section 2 contains the description of observations. In section 3, results and discussion are described, while section 4 presents the Conclusions.

## 2. Observations

For this study we have taken 152 SEPs, which are associated with CMEs and solar flares. These SEPs are taken in the energy channel >10 MeV during solar cycle 23 and 24 from 1997 to 2017. 106 SEPs are reported during solar cycle 23 and during cycle 24, 46 SEPs are reported in this energy channel. The SEP intensity is given in the unit of particle flux unit (pfu) and 1 pfu = 1 proton $cm^{-2}s^{-1}sr^{-1}$. The SEP list is downloaded from Coordinated Data Analysis Workshops (CDAW) data center (https://cdaw.gsfc.nasa.gov/CME_list/sepe/) [34, 51]. Then each and every event is cross checked with the help of CME data taken from the Large Angle and Spectrometric Coronagraph (LASCO) [52] onboard Solar and Heliospheric Observatory (SOHO) satellite.

Association of each SEP event with CMEs and X-ray flares is cross verified with the help of the movies given in the LASCO Catalogue for each CME event (e.g., https://cdaw.gsfc.nasa.gov/movie/make_javamovie.php?date=19971104&img1=lasc2rdf). GOES X-ray data, which can be found at https://satdat.ngdc.noaa.gov/sem/goes/data/plots/ , is used to cross-verify SEPs that are not included in the LASCO catalogue (CME events are absent for these durations).

The CME source location is taken again using images and movies available at SOHO/LASCO CME catalogue (soho/lascocme catalogue/java movies/). We have carefully checked the source location and active regions (ARs) from GOES soft X-ray flare data. Corresponding to the each active region we have taken the Hale class magnetic classification [53, 54] of the magnetic complexity from https://www.swpc.noaa.gov/products/solar-region-summary and www.solarmonitor.org. According to original Hale classification, designation (α) is for those ARs which have unipolar sunspot or sunspot group. If two sunspots or sunspot groups are of opposite polarities in a region, that active region is assigned as (β). Hale class (γ) is defined as the AR in which positive and negative polarities are distributed in a very complex way, these can't be separated from each other to classify them as a bipolar sunspot group. In 1965, Kunzel [62] proposed a new classification of ARs in addition to existing Hale classes, which was named as (δ) class of ARs. In this classification within a single penumbra separated by no more than 2° in heliographic distance (24 Mm or 33″ at disk center), at least one sunspot contains opposite magnetic polarity [54]. Gradual and impulsive SEP events are extracted from the available list of SEPs. In the given list of major SEP event data the end time of the events is not available. Several researchers calculated SEP end times for their studies on SEP duration with some solar activity features [55-57]. But they didn't list the end times and subsequently the duration of SEPs. Firoj et al. 2022 [58] did a study on duration and fluence of major SEP events and listed the end times of SEPs but they had used only 34 major SEP events.

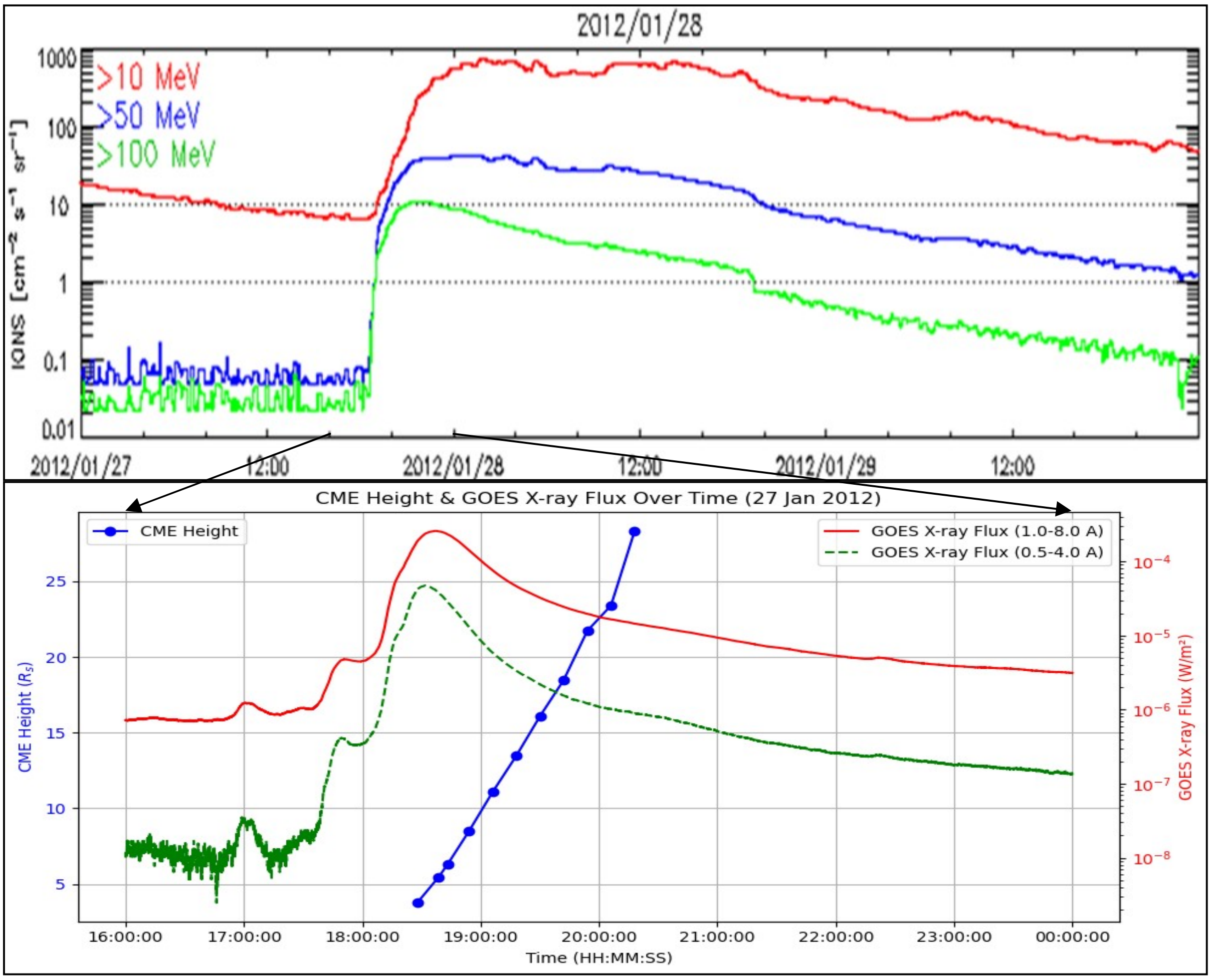

**Figure 1** *An example of gradual SEP event in the energy channel > 10 MeV on 27 January, 2012 with associated CME and solar X-ray flare.* Top panel: Temporal evolution of proton flux. Bottom panel: Corresponding GOES flare temporal evolution in 1-8 A (red colour) and 0.5-4 A (green colour) overlaid by associated CME (blue colour).

We have calculated the SEP duration for 148 SEP events. Four events are corresponding to those events for which end time was not shown because of the unavailability of the CME data on subsequent days. We have determined the duration by noting the onset time of each event and end phase is considered at the time when the event is decaying to the equivalent background intensity level of onset time. Some of the events are overlapping events, for such events end time is the onset time of the next event and at the end of the last overlapped event, end phase is taken at the time when it reaches the background intensity level.

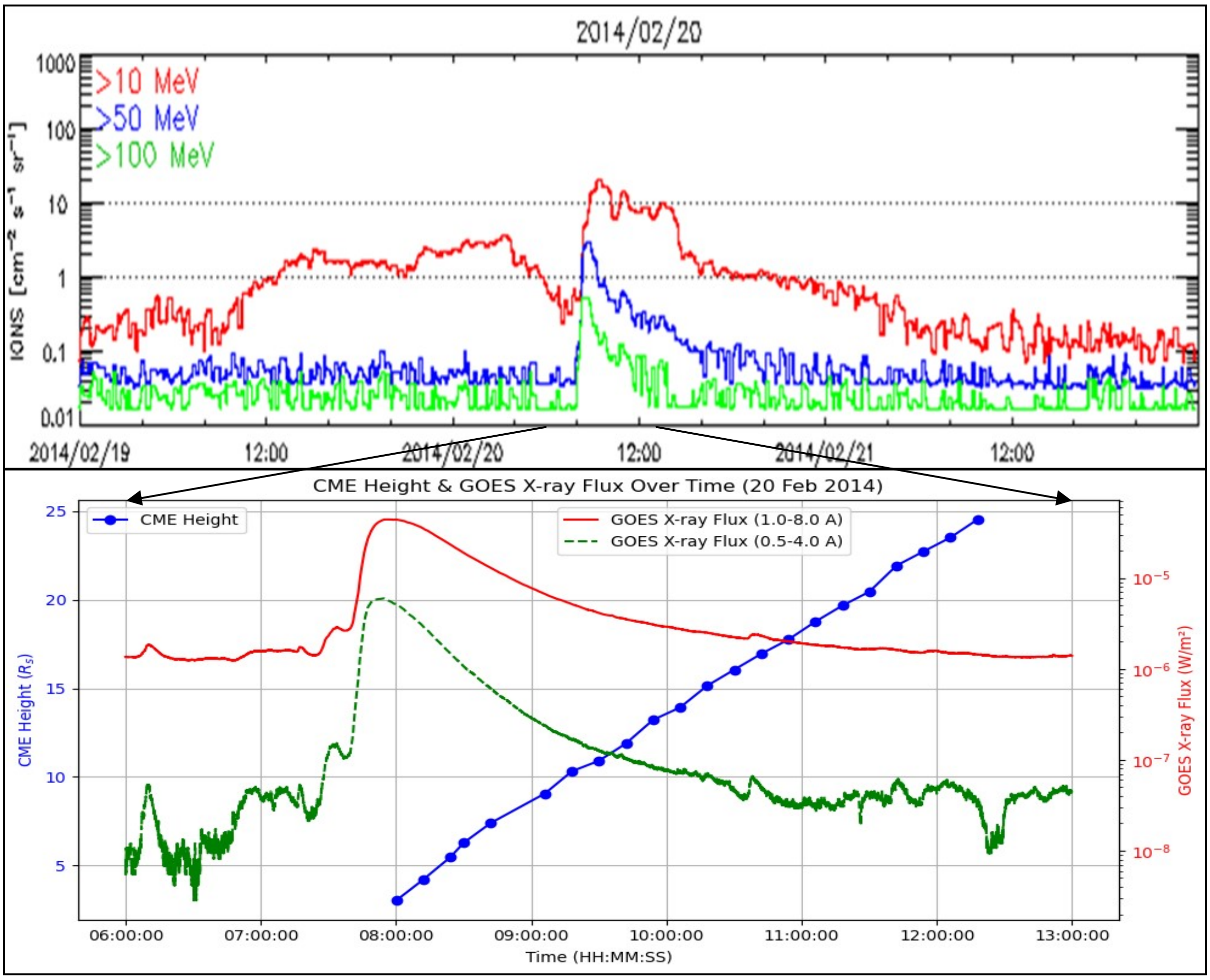

**Figure 2** *Example of impulsive SEP event of 20 February, 2014. Other information is same as described in figure 1.*

In Figure 1, an example of a gradual SEP event is presented, this event has the SEP duration of 4.58 days with the start time of 18:55 UTC on 27-01-2012 and end time of 8:52 UTC on 01-02-2012. This event was associated with a halo CME having a speed of 2508 kms$^{-1}$ and accompanying X-ray is a X1.7 GOES class flare with the source region having βγ magnetic configuration. Figure 2 presents the example of an impulsive SEP event with the SEP duration of 0.8 days with the start time of 8:15 on 20-02-2014 and event was completed at 3:33 UTC of 21-02-2014. This SEP event is associated with a halo CME having speed of 948 kms$^{-1}$ and related X-ray flare has a M3.0-class flare with the source region having complexity α. The source location of the mentioned SEP event is S15W73. The detailed list for the SEP duration of these events is provided here for the future use in Appendix as Table 2.

## 3. Results and Discussion

Different statistical results corresponding to the analysis of SEPs with different activity features like SEP duration, solar X-ray flares, CME properties and magnetic complexity of active regions are presented in following sections.

### 3.1 Association of Flare Properties with SEPs and Hemispheric Distribution

We have calculated the duration of each SEP event and classified them in impulsive and gradual events according to the definition given in the introduction section. The duration of SEPs is plotted against SEP intensity in the Figure 3(a).

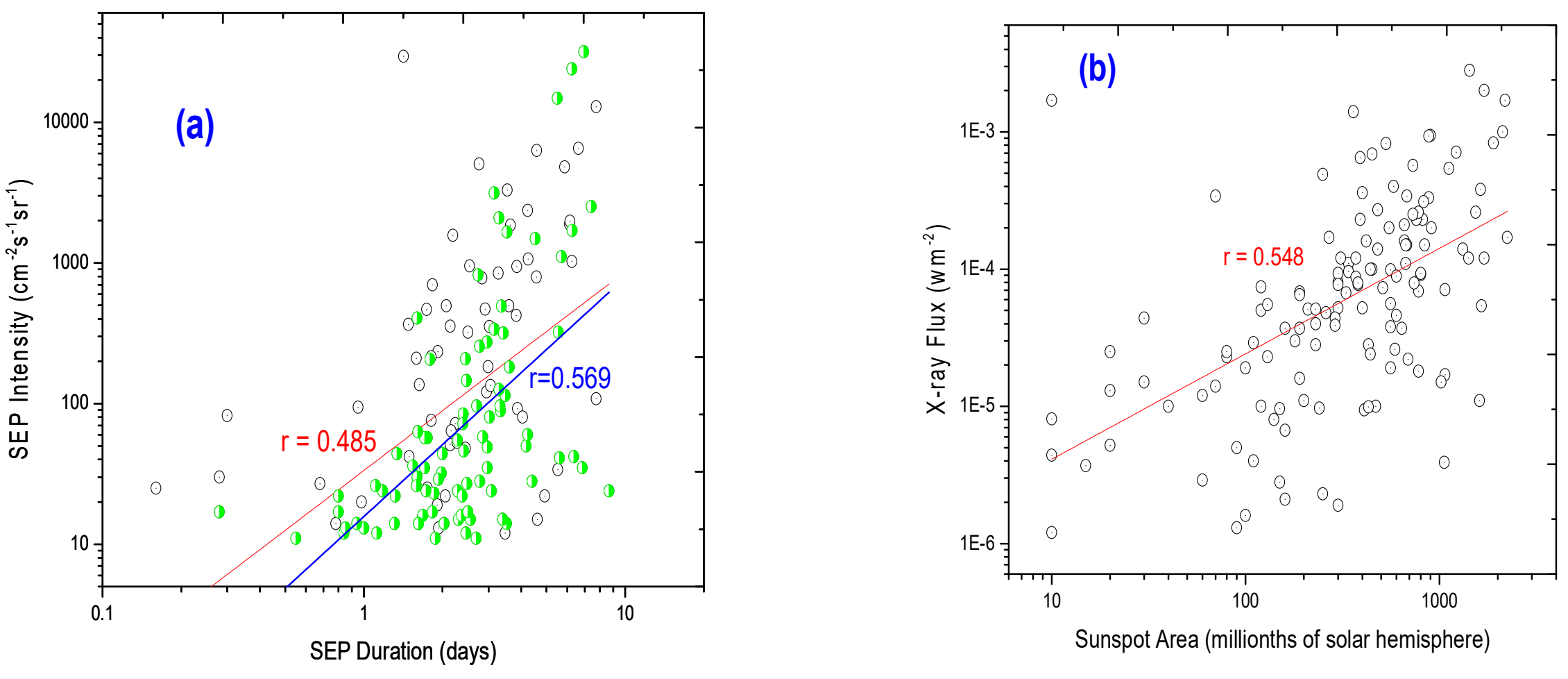

**Figure 3** *Correlation study between SEP intensity and SEP duration (panel-a) taking all the events in to account (red colored line and black balls) and removing overlapping events (blue colored line and green balls) and between X-ray flux and Sunspot area (total sunspot area present in the active region) (panel-b).*

For a scattered plot, the Pearson's correlation coefficient 'r' is a measure of how close the observations are to a line of best fit. The Pearson's correlation coefficient is given by

$$r = \frac{n\sum xy - \sum x \sum y}{\sqrt{[n\sum x^2 - (\sum x)^2][n\sum y^2 - (\sum y)^2]}} \quad (1)$$

Where, n is the total number of x or y variables. Interchanging the variables x and y between two related parameters don't affect the correlation coefficient.

The correlation between SEP intensity and SEP duration is 0.49 for all the events taken in to account. Further for the better association between these parameters, we have inspected all the events and found 35 overlapping SEPs. Now the correlation becomes stronger and the value is 0.57 (Figure 3(a)). In both the cases correlation is moderate; we can infer from this study that gradual nature of the SEPs is not only shown by increment in SEP intensity but by other parameters also like strong CME properties. Plotting the flare X-ray flux against sunspot area *(total sunspot area present in the active region)* (Figure 3(b)) yields a correlation coefficient of 0.55, suggesting a moderate correlation between the two. The X-ray flux is the flux measured in 1-8 Å wavelengths in the unit of $Wm^{-2}$. This full disk X-ray flux is measured by GOES satellite assuming sun as a star. It means active regions with larger sunspot areas generally generate larger flares. Larger will be the sunspot area more will be the magnetic flux so higher will be the probability of generating strong flares [42, 50]. However, the flare generation depends more on the magnetic complexity of the active regions [47, 59]. This study of magnetic complexity of ARs is described in section 3.2.

In Figure 4(a), we have plotted the distribution of SEP events in different hemispheres of the solar disk. We found that the latitudinal distribution varies between $-40^0$ to $+ 40^0$, while longitudinal distribution spans the whole longitude range ($-90^0$ to $+90^0$). We found approximately equal distribution of SEP events in northern and southern hemispheres (71 in north and 68 in south). But we see a big difference in the distribution of events between eastern and western longitudes (31 in east and 114 in west), which is because of Parker's spiral lines [29, 30, 37]. Our analysis shows that although CME speeds and X-ray fluxes are higher in the eastern hemisphere, the average SEP intensity beyond 40° east is actually lower compared to that in the western hemisphere beyond 40° west.

80% of the events are located in the western hemisphere and only 20% are in the eastern hemisphere. In Figure 4(b), a time series of the latitudinal distribution is displayed for both the cycles. We found a trend of butterfly diagram for both the cycles i.e., migration of events from higher latitudes (here $35^0$) towards equator ($1^0$) and a shape of wings of butterfly is obtained. The SEP events are originated at the higher latitudes (both northern and southern hemispheres) at the beginning of the cycle and continuously shift towards the center of the disk as the cycle progresses.

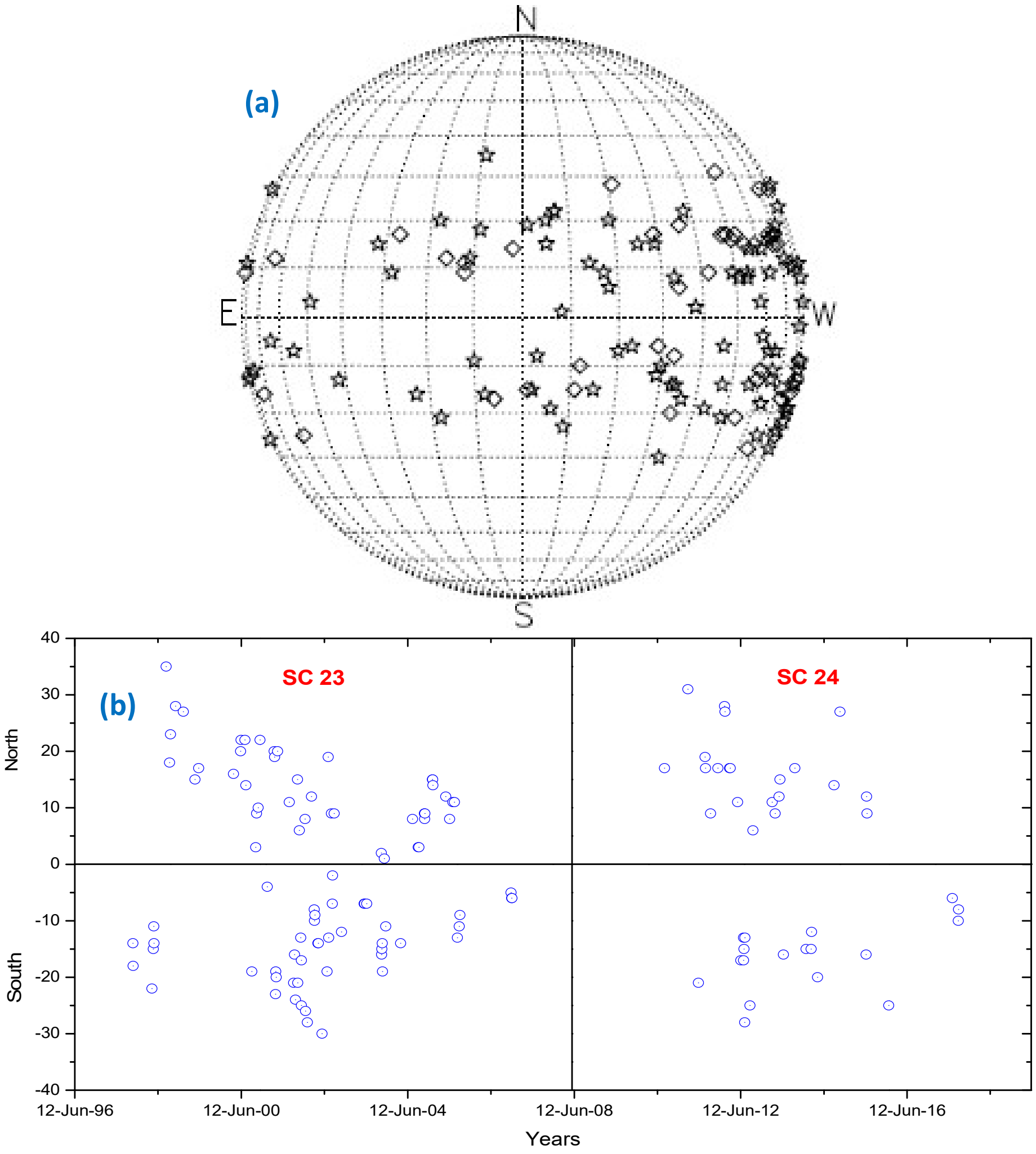

**Figure 4** *Source location of SEP events on solar disk (panel-a). Locations with asterisk (*)are showing the events of cycle 23 and those with diamond (◊) showing the events of cycle 24. Latitudinal time series distribution of SEP events is shown in panel - b.*

A correlation analysis between SEP intensity and X-ray flux was conducted for solar cycles 23 and 24, which is presented in Figure 5 (a). We obtained a correlation coefficient of 0.46, indicating a moderate relationship between the two variables. Gopalswamy et al., 2003 and 2004 [30, 37] also reported a poor correlation of 0.41 between SEPs intensity and X-ray flare flux.

Our study of the correlation between SEP intensity and various X-ray flare classes (Figure 5 (b, c, d)) shows that the correlation strengthens as flare intensity increases, from weaker (C class) to stronger flares (X class). The correlation between C-class flares and SEP intensity is 0.06 only, while that between M-class and SEP intensity becomes 0.24. This correlation increases to 0.37 when we consider X-class flares with SEP intensity. The negative correlation between flare C-class and SEPs intensity is 0.06 (close to zero), which shows no relation between them. The reason for obtaining moderate correlation between these two parameters can be because of the association of SEP events not only with solar flares but with other phenomena also like magnetic topology of active regions and shock wave generated CMEs.

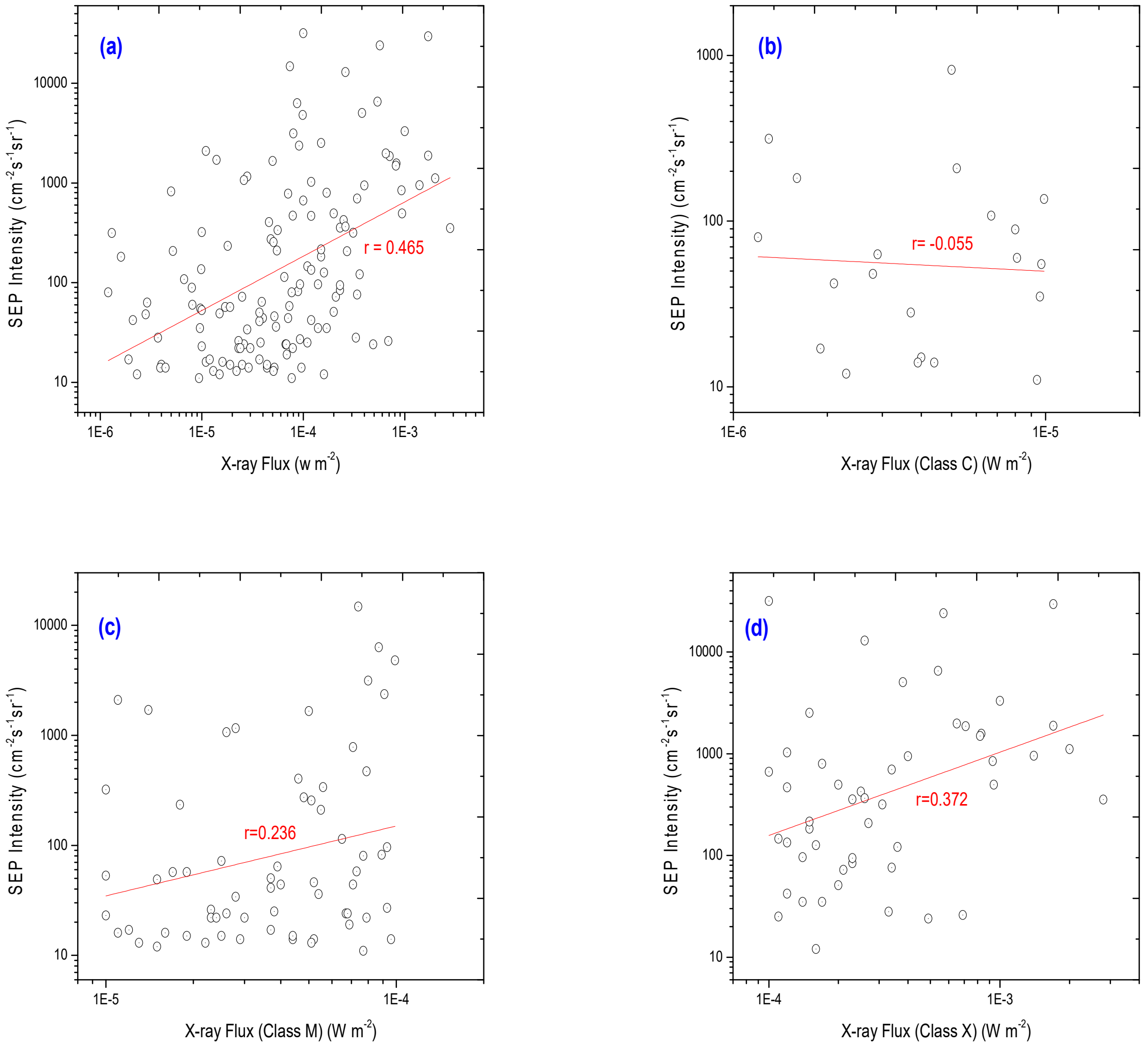

**Figure 5** *Variation of total X-ray flux with SEP intensity and class wise variation of the same with SEP intensity.*

In Figure 6, the SEP duration is plotted against X-ray flux. In Figure 6(a), all the events are considered. We found a weak correlation between X-ray flux and SEP duration which is just 0.26. But when we considered only impulsive events for this study (Figure 6 (b)) we found a negative weak correlation (-0.21) between them. We removed the outlier event lying at the right top corner of the Figure 6(b) and found that the negative correlation is increased to -0.40.

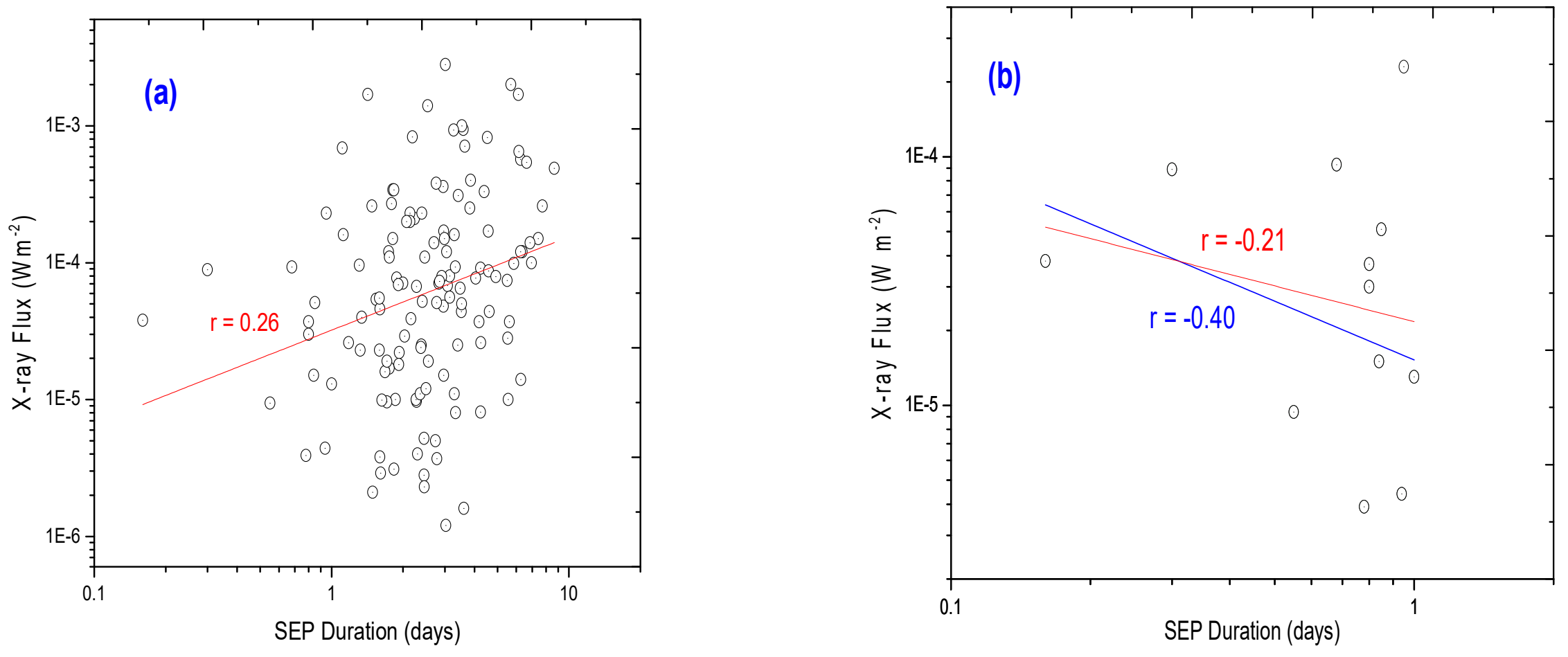

**Figure 6** *Scatter plots showing correlation study of SEP duration with X-ray flux including all events (panel-a) and only impulsive SEP events (panel-b).*

We also checked this correlation study for gradual events only but the correlation was decreased to 0.25. The SEP events under investigation are solar proton events and mostly associated with gradual SEP events [13, 14]. So, more impulsive events are associated with stronger GOES X-ray flux. This is because of the more and impulsive energy released during the impulsive events as well as increment in duration reflects decrement in impulsiveness. Therefore we are getting a negative correlation between them.

### 3.2 SEPs and Magnetic Complexity of Source Regions

Figure 7(a) shows the variation of magnetic configuration of associated active regions. Most of the SEP events (65 events) are associated with the active regions having the most complex magnetic configuration i.e., Hale class βγδ. 31 events are associated with the active regions with magnetic complexity of β configuration. With the active region βγ, the associated number of events is 28. Very few events are associated with the rest of the magnetic complexities (8 with α, 4 with βδ, 3 with αγ and single-single events with αγδ and γδ).

Plotting the average SEP intensity against magnetic complexity classes (Figure 7(b)) revealed that βγδ active regions exhibited the highest average intensity, reaching 2051 pfu. It means the large SEP events are associated with these active regions having βγδ class of magnetic complexity. Since only 4 events were associated with active regions βδ, which contains one event with SEP intensity 5040 pfu. This explains why the average value comes out to be 1982 pfu. Out of these 4 events the last three events are overlapping events. This overlapping event started on 16/01/2005 and continued on 17/01/2005 as second overlapping event with SEP intensity 5040 pfu and ended in 20/01/2005 as third overlapping event. For the active regions with β magnetic complexity and βγ complexity, associated average SEP intensity is 754 pfu and 556 pfu respectively. Compared to the active region β, the average SEP intensity linked to the active region α is significantly lower.

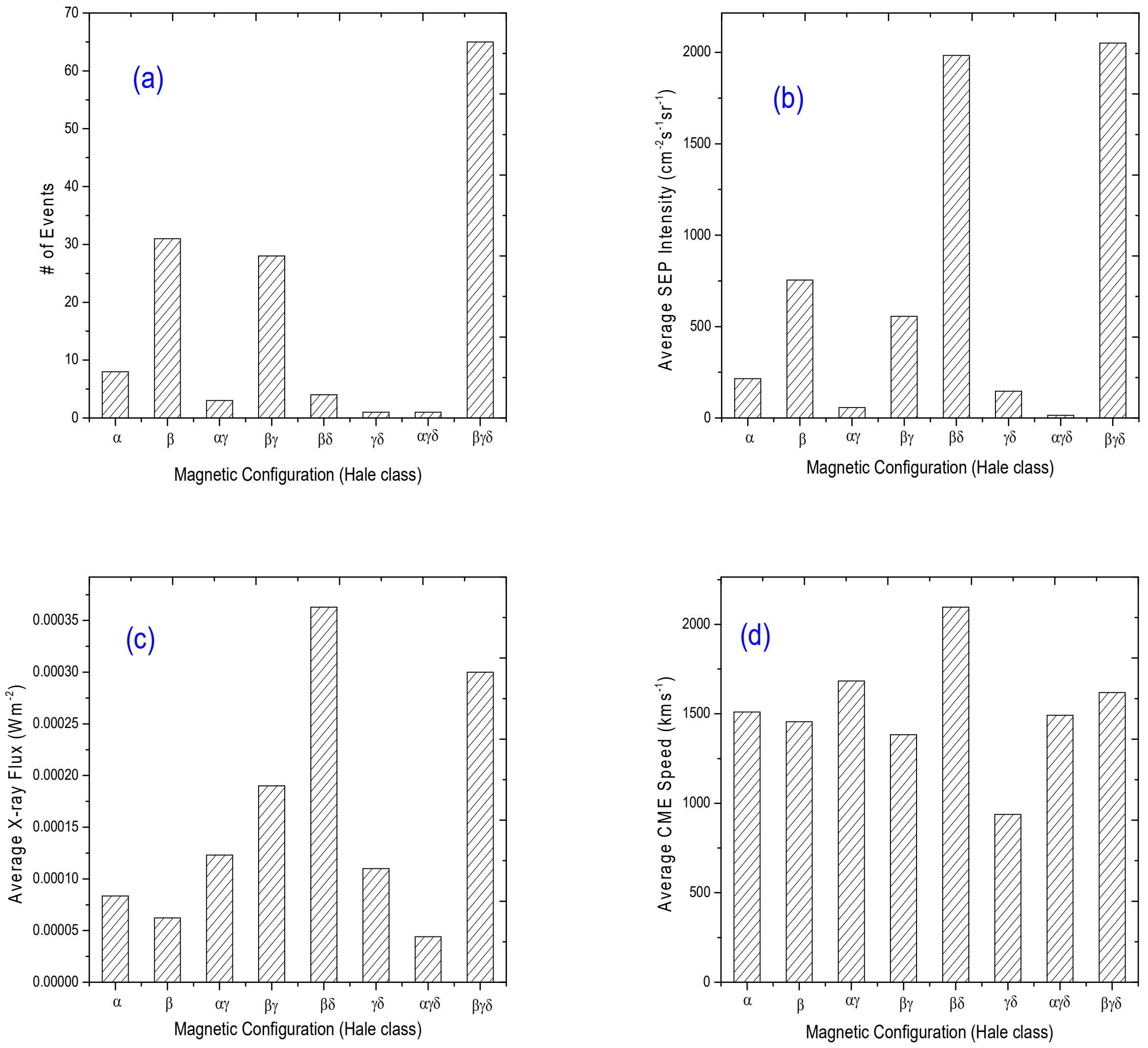

**Figure 7** *Histograms showing the distribution of magnetic configuration (Hale class) with number of SEP events (panel-a), average SEP intensity (panel-b), average X-ray flux (panel-c) and average CME speed (panel-d).*

Similar results are found for the X-ray flux also (Figure 7(c)). Maximum average flux is found for the associated active regions βγδ and βδ (3.00 x $10^{-4}$ and 3.63 x $10^{-4}$ $Wm^{-2}$ respectively). So we find that those active regions which are magnetically more complex are responsible for the generation of high energetic solar flares.

Upon examining the histogram of magnetic complexity of active regions versus average CME speed (Figure 7(d)), we found that the highest average speed corresponds to active regions of the Hale class βδ. This average speed is 2096 $kms^{-1}$ and average speed of 1618 $kms^{-1}$ is found for the magnetic configuration of βγδ. For the active regions having magnetic configuration of Hale class αγ, the average speed is 1683 $kms^{-1}$. Minimum average speed is 938 $kms^{-1}$, which is associated with active regions γδ, while minimum speed of listed CME is 296 $kms^{-1}$. Two faster CMEs (2684 and 3163 $kms^{-1}$) are associated with very large active regions having complexity βγδ. Nine faster CMEs having speed greater than 2500 $kms^{-1}$ are also associated with active regions having very complex magnetic configuration namely βγδ, βγ and βδ. This again shows the capability of large complex active regions to generate the faster CMEs as reported by Michalek and Yashiro, 2013 [29].

**3.3 Association of SEPs with CME properties**

Average CME acceleration and deceleration was plotted with magnetic complexity of solar active regions (Figure 8(a)), we found that 58 SEP events are accelerated and 82 are decelerated. Average CME acceleration is highest again for most complex Hale class of magnetic configuration, which are βγδ and αγδ. Highest average deceleration was found corresponding to the active region βδ. The maximum deceleration recorded was 143 $ms^{-2}$, whereas the highest CME acceleration observed was 83 $ms^{-2}$. All the events associated with active regions αγ and αγδ are accelerated, while all the events corresponding to active regions γδ are decelerated.

Average CME width was also plotted with magnetic complexity of solar active regions (Figure 8(b)), we found that highest average CME width was found for the magnetically most complex active regions βδ, γδ, αγδ, and the value is $360^0$, which means all the events associated with these active regions are also associated with halo CMEs. The average CME width corresponding to active region βγδ is 346, which is also close to halo CMEs. This implies that larger (CME width) CMEs are mostly originated from magnetically very complex active regions.

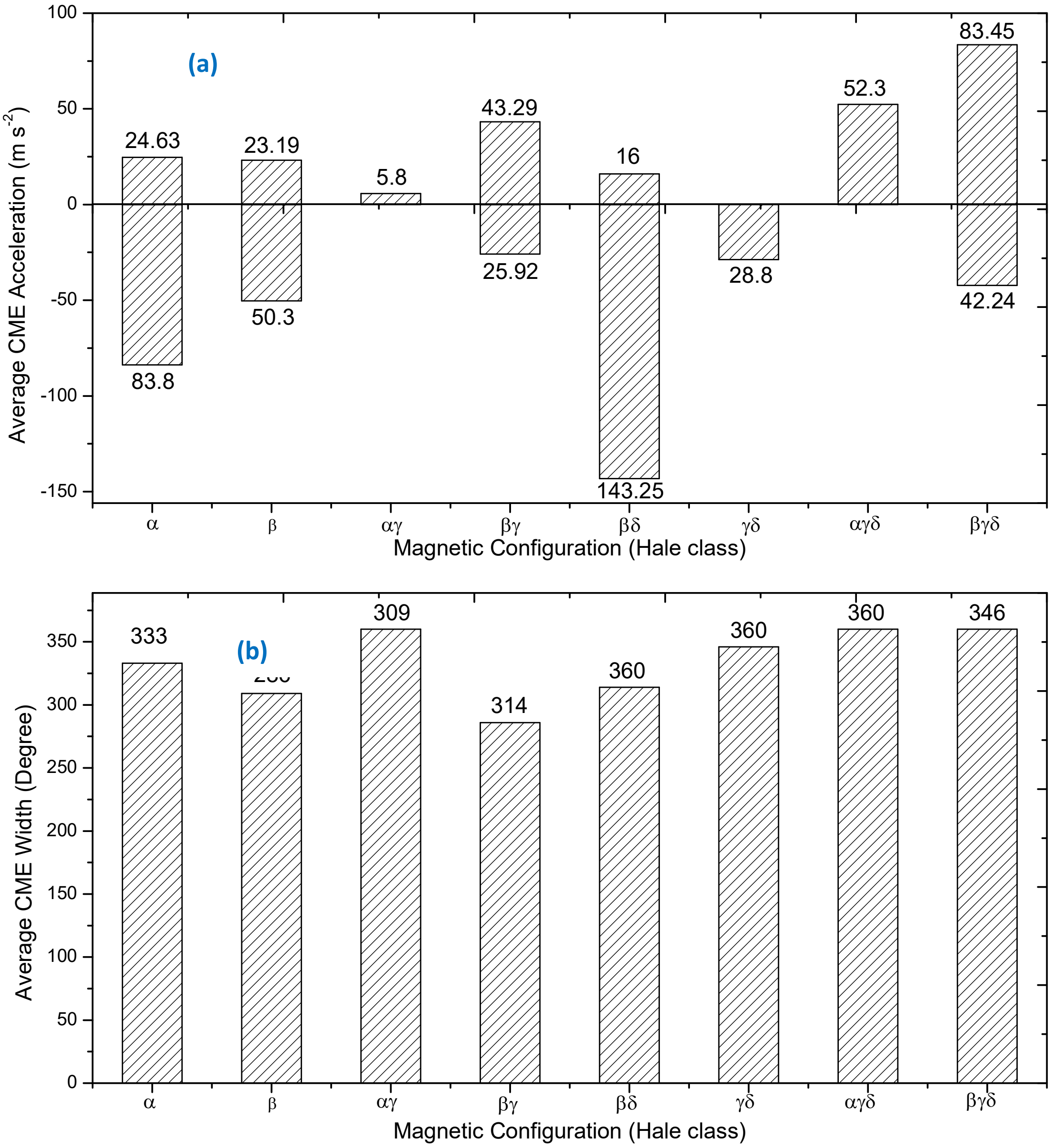

**Figure 8** *Histogram of average CME acceleration and deceleration with magnetic configuration (a) and histogram of average CME width with magnetic configuration (b).*

All the CMEs associated with impulsive SEP events are decelerated except two CMEs. Average CME Acceleration for impulsive events was -17.97 $ms^{-2}$, while that for the gradual SEP events was -4.37 $ms^{-2}$. The average CME speed associated with impulsive SEP events was 1125 km $s^{-1}$ and 1561 $kms^{-1}$ for the gradual events. For the gradual events, the associated CMEs are having lager width ($327^0$) than the associated CME width ($262^0$) for impulsive events.

### 3.4 Gradual and Impulsive events and Comparative study of Solar cycle 23 & 24

By determining the duration of each SEP event, we identified the impulsive and gradual events. The result is plotted in the panel (a) of Figure 9. We found only 10% impulsive events which are characterized as the SEP events having SEP duration of one day or less than that. Additionally, 90% of events are gradual, with SEP duration longer than a day. Although thousands of impulsive events are observed per year and few tens of gradual events are observed in a year [14-16], but we only looked at proton-rich events in our study, not electron-rich ones, which is why we found very few impulsive events.

We found that all the impulsive SEP events are originated from western hemisphere which is in well agreement with previous findings [13, 14]. We observed that all the impulsive events have SEP intensity less than 100 pfu. Many of the events are associated with C class flares or B class flares (very weak flares). We can infer from this observation that impulsive events are also associated with other solar activity features. According to Reams, 2021 [60], impulsive SEPs are also associated with magnetic reconnection in solar jets and jets with fast narrow CMEs.

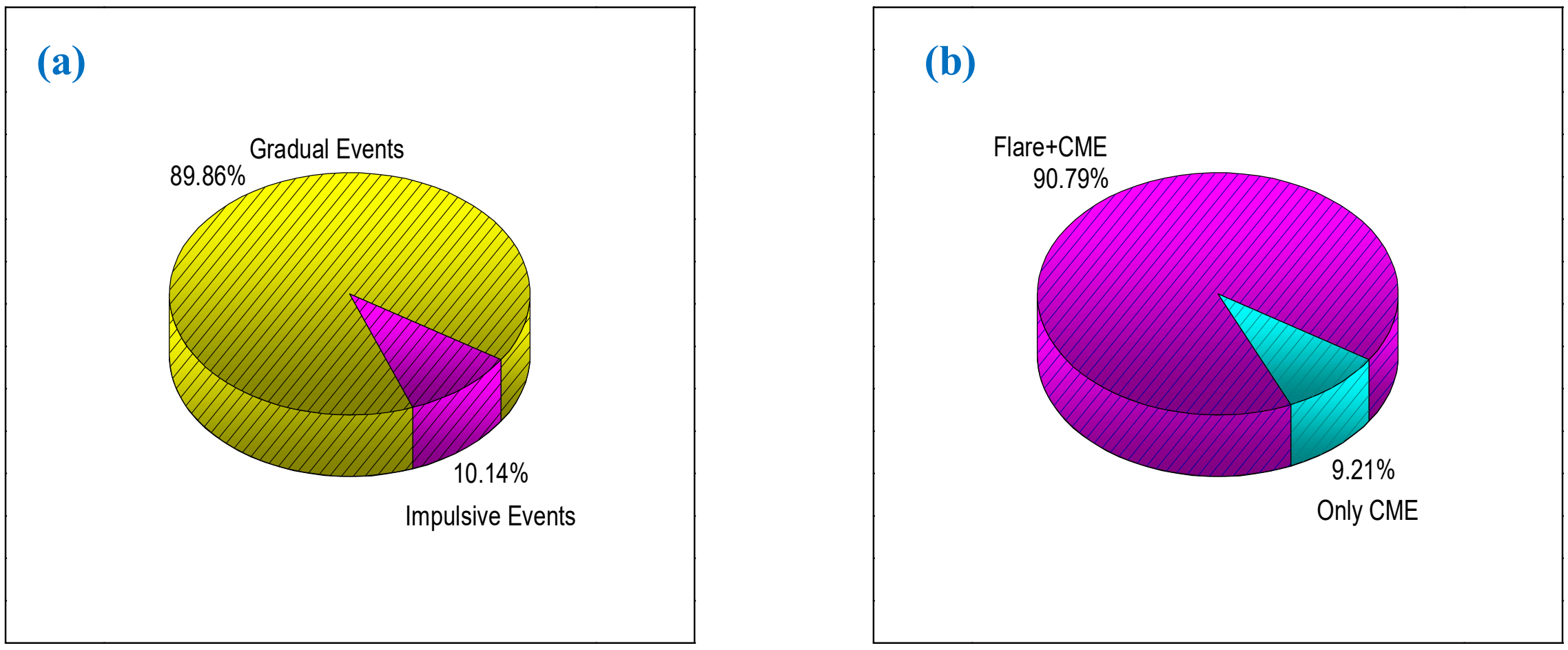

**Figure 9** *Pie diagram showing the gradual versus impulsive SEP events (panel- a) and pie diagram showing events associated with flares and CMEs both versus events associated with only CMEs or CMEs and very weak flares(GOES C-class or weaker than that i.e., B-class) (panel-b).*

We have characterized 10% SEP events which are associated with only CMEs or CMEs and having very weak flares i.e., C-class or B-class flares (Figure 9 (b)). We observed that maximum of these events are associated with halo CMEs and all the events are located at the limb of western hemisphere of the solar disk, only one is from the eastern limb of the solar disk. Two impulsive SEP events are there to which the flare association is very weak. This may be because these events might be associated with solar jets also.

Solar cycles 23 and 24 are compared in table 1 to have a clear view between these two cycles. In solar cycle 23, a total of 106 SEP events were recorded, compared to only 46 events in cycle 24. Halo CMEs accounted for 75% of events in cycle 23, whereas a higher percentage (86%) was observed in cycle 24. The average SEP intensity during cycle 23 was 1524 pfu, which significantly decreased to 485 pfu in cycle 24. Similarly, the average X-ray flux dropped from $2.33 \times 10^{-4}$ $Wm^{-2}$ in cycle 23 to $1.3733 \times 10^{-4}$ $Wm^{-2}$ in cycle 24.

**Table 1** Comparison between solar cycle 23 and 24.

| **Different Activity Features** | **Cycle 23** | **Cycle 24** | **Total** |
|---|---|---|---|
| Number of SEP Events | 106 | 46 | 152 |
| No. of Halo CMEs | 73 | 40 | 113 |
| Average SEP Intensity | 1524 pfu | 485 pfu | 1210 pfu |
| Average X-ray Flux | $2.33 \times 10^{-4}$ $Wm^{-2}$ | $1.37 \times 10^{-4}$ $Wm^{-2}$ | $2.05 \times 10^{-4}$ $Wm^{-2}$ |
| Average CME Speed | 1521 $kms^{-1}$ | 1510$kms^{-1}$ | 1518 $kms^{-1}$ |
| No. of Events in Eastern Hemisphere | 20 | 10 | 30 |
| No. of Events in Western Hemisphere | 80 | 36 | 116 |
| No. of Events in Northern Hemisphere | 50 | 23 | 71 |
| No. of Events in Southern Hemisphere | 50 | 23 | 68 |

Average CME speed in cycle 23 is 1521 $kms^{-1}$ and it is decreased to 1510 $kms^{-1}$ for cycle 24. Number of SEP events in east, west, north and southern hemispheres is 20, 80, 50 and 50 respectively in case of cycle 23, while these numbers become 10, 36, 21 and 18 respectively for cycle 24. Above results clearly decide that solar cycle 24 is a weak cycle compared to cycle 23. Gopalswamy et al., 2022 [63] also revealed that sunspot numbers, high energy SEPs are less in solar cycle 24, photospheric magnetic field is also weaker in this cycle. This comparative study of solar cycle 23 and 24 can be combined with the similar kind of study for previous solar cycle. Finally these results can be used for the prediction of future solar cycle strength using the latest Artificial Intelligence technique.

## 4. Conclusions

In this study, we have investigated 152 major SEP events associated with solar X-ray flares, CME parameters, active region characteristics and SEP durations. For this investigation, a wide range of data was collected during solar cycles 23 and 24. A summary of the study is given below.

Solar cycles 23 and 24 were compared, and it was found that cycle 24 was weaker. The average SEP intensity was 1210 pfu, indicating that most SEPs were intense. Only a few SEPs had low intensity, and these were mostly associated with impulsive events. With a 7-day SEP duration, the highest SEP intensity recorded in the dataset was 31,700 pfu, originating from an active region with the magnetic configuration βγδ. The linked flare belongs to the X class, and its associated CME speed exceeds 1000 $kms^{-1}$. The same pattern is seen in events with intensities more than 10,000 pfu, supporting previous findings that strong SEPs are caused by strong flares, large CMEs, and magnetically complex active regions.

According to a recent study by Marroquin, R., D., et al. 2023 [61], the active region β magnetic configuration has generated the greatest number of SEP events. The majority of the SEPs are linked to active regions βγδ and β in our study; very few events are associated with unipolar active regions. This difference may be because of the chosen SEP events. We have taken only 20 year data in an energy channel > 10 MeV and all the SEPs are having intensity ≥10 pfu. Moreover an adequate number of SEPs are originated from the western limb of the Sun, where information about the active regions may not be reliable due to projection effects.

The average SEP intensity for the βγδ active region class is 2051 pfu, the highest average intensity for this configuration when compared to other magnetic active region configurations. The events that are associated to active regions βγδ and βδ have the highest average X-ray flux. It indicates that the active regions with the most complex magnetic configuration are the ones that produce strong flares. 80% of events originate in the western hemisphere, which is consistent with earlier findings resulting from Parker's spiraling. The magnetic complexity of active regions is used to study the CME properties. We discovered that two of the faster CMEs, with respective speeds of 2684 and 3163 $kms^{-1}$, are associated to extremely large active regions βγδ. The nine most rapid CMEs (with speeds above 2500 $kms^{-1}$) are likewise linked to active regions with extremely complex magnetic configurations βγδ, βγ, and βδ. Therefore, we may conclude that the production of high speed CMEs is caused by the magnetically complex active regions. The magnetically most complicated active regions βδ, γδ, αγδ, and βγδ have the highest average CME width, indicating that large CMEs are a result of these active regions.

Only 10% of the major SEP events in our data set were impulsive; the remaining 90% were gradual. Since many of these impulsive events are associated with C class or very weak flares and their SEP intensity is less than 100 pfu, which indicate that impulsive events are not only caused by solar flares but also by other activity features like solar jets. The western hemisphere of the solar disk is the source of all impulsive events.

In the data set used for this study, the maximum SEP intensity is characterized as 31700 pfu. This event is a large gradual event having SEP duration of approximately 7 days. This event is, originated from the magnetically most complex active region βγδ [42]. Associated flare is an X-class flare and CME speed is more than 1000 $kms^{-1}$. This associated CME is decelerated with 63.4 $ms^{-2}$. All the 5 events having intensity more than10000 pfu show similar type of behaviour. So from these results we can infer that the most intense SEP event is a consequence of the combined effect of strong X-ray flare, large CME and most complex magnetic topology. Which confirms the earlier findings and predictions that more magnetically complex regions are responsible for the generation of strong flares and CMEs hence the strong SEPs [42, 48-50].

This work leads us to the conclusion that shock-generated CMEs, solar flares associated with magnetic reconnection, and the magnetic complexity of solar active regions all contribute together to the generation of SEP events rather than being caused by a single mechanism. Both solar flares and solar jets are responsible for the impulsive events. Shocks produced by CMEs are primarily linked to gradual SEP events.

**Acknowledgements**

We acknowledge the open data policy of NGDC, SOHO and SDO. This study made use of NASA Astrophysics Data System Bibliographic Services. We are very thankful to reviewers who gave valuable suggestions to improve the quality of paper.

**Statements and Declarations:**

**Funding** – We declare that no funding, grants or other support were received during the preparation of this manuscript.

**Competing Interests -** *The authors have no relevant financial or non-financial interests to disclose.*

***Author Contributions*** *– All authors contributed in this study equally. Material preparation and data collection was performed by Raj Kumar and Ramesh Chandra. Analysis was performed by all the authors viz. Raj Kumar, Ramesh Chandra, Bimal Pande and Seema Pande. First and corresponding author has written the first draft of the manuscript and all the authors commented and contributed to make it final.*

## Appendix: Supplementary Information: Table 2.

| Day | start time hours | End date & time | SEP duration days | SEP (pfu) | CME | | | | | | Flare class | **NOAA Active Region Number** | hale class | **Total Sunspot Area present in the active region** | Number of spots |
|---|---|---|---|---|---|---|---|---|---|---|---|---|---|---|---|
| | | | | | Day | time | Acceleration | speed | width | location | | | | | |
| 04-11-97 | 6.66 | 06-11-97 12.5 h | 2.24 | 72 | 04-11-97 | 6:10 | -22.10 | 785 | H | S14W33 | X2.1/2B | 8100 | βγδ | 660 | 34 |
| 06-11-97 | 12.5 | 10-11-97 2.6h | 3.59 | 495 | 06-11-97 | 12:10 | -44.10 | 1556 | H | S18W63 | X9.4/2B | 8100 | βγδ | 900 | 19 |
| 20-04-98 | 11.25 | 26-04-98 17.77h | 6.27 | 1700 | 20-04-98 | 10:07 | 43.50 | 1863 | >243 | S22W90 | M1.4/-- | (8205)8194 | β | 70 | 12 |
| 02-05-98 | 14 | 05-05-98 1.33 h | 2.47 | 146 | 02-05-98 | 14:06 | -28.80 | 938 | H | S15W15 | X1.1/3B | 8210 | γδ | 340 | 30 |
| 06-05-98 | 8.42 | 08-05-98 3.55 h | 1.79 | 207 | 06-05-98 | 8:29 | 24.50 | 1099 | 190 | S11W65 | X2.7/1N | 8210 | β | 480 | |
| 09-05-98 | 6.33 | 11-05-98 3.55 h | 1.88 | 11 | 09-05-98 | 3:35 | 140.50 | 2331 | 178 | S14W89 | M7.7/-- | 8214 | β | 380 | 21 |
| 24-08-98 | 23.17 | No data | | 665 | No data | | | | | N35E09 | X1.0/3B | 8307 | βδ | 450 | 12 |
| 23-09-98 | 14.33 | 25-09-98 14.22h | 2.00 | 44 | No data | | | | | N18E09 | M7.1/3B | No data | | | |
| 30-09-98 | 14.33 | No data | | 1160 | No data | | | | | N23W81 | M2.8/2N | 8340 | α | 230 | 1 |
| 07-11-98 | 22.42 | 08-11-98 11.55h | 0.55 | 11 | 07-11-98 | 20:54 | 23.70 | 750 | 321 | | C9.4 | 8375 | βγδ | 410 | 41 |
| 14-11-98 | 6.5 | No data | | 314 | No data | | | | | N28W90 | C1.3/-- | 8375(8383) | β | 90 | 10 |
| 20-01-99 | 22.83 | | | 14 | | | | | | N27E90 | M5.2/-- | 8439 | βγ | 300 | 44 |
| 24-04-99 | 15.5 | 26-04-99 15.11h | 1.98 | 32 | 24-04-99 | 13:31 | 37.10 | 1495 | H | >NW90 | | 8516 | β | 30 | 8 |
| 04-05-99 | 3 | 07-05-99 15.55h | 3.52 | 14 | 03-05-99 | 6:06 | 15.80 | 1584 | H | N15E32 | M4.4/2N | 8525/8524 | β | 290 | 16 |
| 01-06-99 | 21 | 04-06-99 08.00h | 2.45 | 48 | 01-06-99 | 19:37 | 1.80 | 1772 | H | >NW90 | C2.8 | 8552 | β | 150 | 11 |
| 04-06-99 | 8.0 | 06-06-99 12.22h | 2.16 | 64 | 04-06-99 | 7:26 | -158.80 | 2230 | 150 | N17W69 | M3.9/2B | 8552 | β | 290 | 5 |
| 18-02-00 | 10.08 | 19-02-00 10 h | 1.00 | 13 | 18-02-00 | 9:54 | -9.60 | 890 | 118 | >NW90 | C1.3/-- | 8872 | β | 20 | 6 |
| 04-04-00 | 17 | 06-04-00 23.77h | 2.28 | 55 | 04-04-00 | 16:32 | 12.80 | 1188 | H | N16W66 | C9.4 | 8933 | β | 240 | 16 |
| 06-06-00 | 22.33 | 09-06-00 08.00h | 2.40 | 84 | 06-06-00 | 15:54 | 1.50 | 1119 | H | N20E18 | X2.3/-- | 9026 | βγδ | 820 | 22 |
| 10-06-00 | 17.66 | 13-06-00 3.55 h | 2.41 | 46 | 10-06-00 | 17:08 | -21.20 | 1108 | H | N22W38 | M5.2/3B | 9026 | βγ | 400 | 33 |
| 14-07-00 | 10.58 | 20-07-00 16.88h | 6.26 | 24000 | 14-07-00 | 10:54 | -96.10 | 1674 | H | N22W07 | X5.7/3B | 9077 | βγδ | 730 | 42 |
| 22-07-00 | 12.42 | 23-07-00 07.55h | 0.80 | 17 | 22-07-00 | 11:54 | -12.40 | 1230 | 105 | N14W56 | M3.7/2N | 9085 | β | 190 | 18 |
| 28-07-00 | 1.75 | 30-07-00 12.88h | 2.46 | 12 | 27-07-00 | 19:54 | -21.40 | 905 | H | >N90 | C2.3 | 9097 | βγ | 250 | 23 |
| 11-08-00 | 14.5 | 11-08-00 21.33h | 0.28 | 17 | 11-08-00 | 1:31 | 4.50 | 296 | 95 | No data | | 9114 | βγ | 300 | 19 |
| 12-09-00 | 13.83 | 18-09-00 3.11 h | 5.55 | 321 | 12-09-00 | 11:54 | 58.20 | 1550 | H | S19W06 | M1.0/2N | 9154 | β | 120 | 5 |
| 16-10-00 | 8.17 | 19-10-00 17.77h | 3.40 | 15 | 16-10-00 | 7:27 | 9.90 | 1336 | H | N03W90 | M2.5/-- | 9192 | β | 20 | 6 |
| 25-10-00 | 13.58 | 27-10-00 20.88h | 2.30 | 15 | 25-10-00 | 8:26 | 17.40 | 770 | H | N09W63 | C4.0/-- | 9198 | β | 110 | 5 |
| 08-11-00 | 23.58 | 14-11-00 11.55h | 5.50 | 14800 | 08-11-00 | 23:06 | 69.90 | 1738 | >170 | N10W77 | M7.4/3F | 9213 | β | 120 | 5 |
| 24-11-00 | 5.32 | 25-11-00 4.00 h | 0.95 | 94 | 24-11-00 | 15:30 | -3.30 | 1245 | H | N22W07 | X2.3/2B | 9236 | β | 390 | 17 |
| 25-11-00 | 4 | 29-11-00 0.44 h | 3.85 | 942 | 26-11-00 | 17:06 | 5.80 | 980 | H | No data | X4.0 | 9236 | βγ | 580 | 29 |
| 28-01-01 | 17.25 | 31-01-01 16.44h | 2.97 | 49 | 28-01-01 | 15:54 | 3.50 | 916 | 250 | S04W59 | M1.5/1N | 9313 | β | 30 | 7 |
| 29-03-01 | 12.25 | 01-04-01 11.55h | 2.97 | 35 | 29-03-01 | 10:26 | 3.50 | 942 | H | N20W19 | X1.7/1F | 9393 | βγδ | 2240 | 51 |
| 02-04-01 | 23.17 | 08-04-01 16.00h | 5.70 | 1110 | 02-04-01 | 22:06 | 108.50 | 2505 | 244 | N19W72 | X20./-- | 9393 | βγδ | 1700 | 53 |
| 10-04-01 | 8.25 | 12-04-01 11.66h | 2.14 | 355 | 10-04-01 | 5:30 | 211.60 | 2411 | H | S23W09 | X2.3/3B | 9415 | βγδ | 760 | 32 |
| 12-04-01 | 11.66 | 14-04-01 15.11h | 2.14 | 51 | 12-04-01 | 10:31 | -20.00 | 1184 | H | S19W43 | X2.0/-- | 9415 | βγδ | 550 | 23 |
| 15-04-01 | 14 | 18-04-01 3.00 h | 2.54 | 951 | 15-04-01 | 14:06 | -35.90 | 1199 | 167 | S20W85 | X14./2B | 9415 | βγ | 360 | 9 |
| 18-04-01 | 3 | 20-04-01 15.11h | 2.50 | 321 | 18-04-01 | 2:30 | -9.50 | 2465 | H | >SW90 | No Flare | 9415 | βγ | 20 | 2 |
| 27-04-01 | 1.08 | 28-04-01 19.11h | 1.75 | 57 | 26-04-01 | 12:30 | 21.10 | 1006 | H | N20W05 | M1.7/-- | 9433 | βγδ | 1070 | 82 |
| 07-05-01 | 14 | 09-05-01 4.44h | 1.60 | 30 | 07-05-01 | 12:06 | 19.20 | 1223 | 205 | >NW90 | C3.8 | 9447 | βγ | 200 | 6 |
| 15-06-01 | 16.42 | 18-06-01 4.00 h | 2.48 | 27 | 15-06-01 | 15:56 | 56.90 | 1701 | H | >SW90 | No Flare | 9487 | βγ | 140 | 3 |
| 09-08-01 | 19.5 | 11-08-01 15.55h | 1.83 | 17 | 09-08-01 | 10:30 | 4.40 | 479 | 175 | N11W14 | C3.1 | 9557 | βγ | 270 | 11 |
| 16-08-01 | 0.75 | 19-08-01 09.33h | 3.36 | 493 | 15-08-01 | 23:54 | -31.70 | 1575 | H | >SW90 | No Flare | 9573 | βγδ | 40 | 5 |
| 15-09-01 | 12.08 | 16-09-01 8.22 h | 0.84 | 12 | 15-09-01 | 11:54 | -4.00 | 478 | 130 | S21W49 | M1.5/1N | 9608 | βγδ | 1020 | 34 |
| 24-09-01 | 11.66 | 01-10-01 12.4 h | 7.74 | 12900 | 24-09-01 | 10:30 | 54.10 | 2402 | H | S16E23 | X2.6/-- | 9632 | βγδ | 780 | 17 |
| 01-10-01 | 12.25 | 05-10-01 18 h | 4.24 | 2360 | 01-10-01 | 5:30 | 97.80 | 1405 | H | S24W81 | M9.1/-- | 9628 | βγδ | 800 | 20 |
| 19-10-01 | 18 | 20-10-01 20.88h | 1.12 | 12 | 19-10-01 | 16:50 | -0.70 | 901 | H | N15W29 | X1.6/2B | 9661 | βγδ | 660 | 24 |
| 22-10-01 | 16.75 | 24-10-01 23.55h | 2.28 | 24 | 22-10-01 | 15:06 | -8.00 | 1336 | H | S21E18 | M6.7/2N | 9672 | βγδ | 330 | 14 |
| 04-11-01 | 16.66 | 11-11-01 15.55h | 6.95 | 31700 | 04-11-01 | 16:35 | -63.40 | 1810 | H | N06W18 | X1.0/3B | 9684 | βγδ | 440 | 27 |
| 17-11-01 | 8.83 | 22-11-01 21.4 h | 5.52 | 34 | 17-11-01 | 5:30 | -22.50 | 1379 | H | S13E42 | M2.8/1N | 9704 | βγδ | 430 | 14 |
| 22-11-01 | 21.25 | 23-11-01 1.083h | 0.16 | 25 | 22-11-01 | 20:30 | -43.30 | 1443 | H | S25W67 | M3.8/2B | 9704 | βγδ | 560 | 30 |
| 23-11-01 | 1.08 | 28-11-01 21.77h | 5.86 | 4800 | 22-11-01 | 23:30 | -12.90 | 1437 | H | S17W36 | M9.9/-- | 9710 | βγδ | 560 | 30 |
| 26-12-01 | 5.83 | 29-12-01 1.83h | 2.83 | 780 | 26-12-01 | 5:30 | -39.90 | 1446 | 212 | N08W54 | M7.1/1B | 9742 | βγ | 1070 | 49 |
| 29-12-01 | 1.83 | 30-12-01 21.42h | 1.81 | 76 | 28-12-01 | 20:30 | 6.90 | 2216 | H | S26E90 | X3.4/-- | 9756/9767 | αγ | 70 | 1 |
| 30-12-01 | 21.42 | 07-01-02 15.11h | 7.74 | 108 | 30-12-01 | 23.3 | -10.20 | 457 | H | No data | | 9754 | βγ | 160 | 28 |
| 10-01-02 | 8.66 | 14-01-02 5.583h | 3.87 | 92 | 08-01-02 | 17:54 | 81.40 | 1794 | H | >NE90 | No data | 9778 | βγδ | 170 | 9 |
| 14-01-02 | 5.58 | 18-01-02 20.44h | 4.62 | 15 | 14-01-02 | 5:35 | 52.30 | 1492 | H | S28W83 | M4.4/-- | 9772 | αγδ | 30 | 1 |
| 20-02-02 | 6.75 | 21-02-02 3.11h | 0.85 | 13 | 20-02-02 | 6:30 | -17.10 | 952 | H | N12W72 | M5.1/1N | 9825 | βγ | 210 | 11 |
| 16-03-02 | 8.42 | 18-03-02 6.75 h | 1.93 | 13 | 15-03-02 | 23:06 | -17.40 | 957 | H | S08W03 | M2.2/1F | 9866 | blank | 690 | 26 |
| 18-03-02 | 6.75 | 20-03-02 13.33h | 2.27 | 53 | 18-03-02 | 2:54 | -2.90 | 989 | H | S10W30 | M1.0/-- | 9866 | βγδ | 470 | 24 |
| 20-03-02 | 13.33 | 21-03-02 12.88h | 0.98 | 20 | 20-03-02 | 14:54 | -5.89 | 439.5 | 27 | No data | | 9866 | βγδ | 290 | 10 |
| 22-03-02 | 13.83 | 24-04-02 6.22h | 1.68 | 16 | 22-03-02 | 11:06 | -22.50 | 1750 | H | S09W90 | M1.6/-- | 9866 | βγδ | 190 | 1 |
| 17-04-02 | 11.58 | 18-04-02 16.0h | 1.18 | 24 | 17-04-02 | 8:26 | -19.80 | 1240 | H | S14W34 | M2.6/2N | 9906 | βγδ | 590 | 27 |
| 21-04-02 | 1.75 | 28-04-02 12h | 7.43 | 2520 | 21-04-02 | 1:27 | -1.40 | 2393 | H | S14W84 | X1.5/1F | 9906 | βγ | 680 | 27 |
| 22-05-02 | 7.17 | 25-05-02 0.88 h | 2.74 | 820 | 22-05-02 | 3:50 | -10.40 | 1557 | H | S30W34 | C5.0/-- | 9954 | βγδ | 90 | 8 |
| 07-07-02 | 12.75 | 09-07-02 9.33h | 1.86 | 23 | 07-07-02 | 11:30 | 22.00 | 1423 | >228 | S19W90 | M1.0/-- | 17 | β | 40 | 2 |
| 16-07-02 | 10.25 | 19-07-02 8.33 h | 1.92 | 234 | 15-07-02 | 21:30 | -7.30 | 1300 | >188 | N19W01 | M1.8/-- | 30 | βγδ | 780 | 71 |
| 19-07-02 | 8.33 | 20-07-02 3.11 h | 0.78 | 14 | No Data | | | Several CMEs | No Data | | c3.9 | 30 | βγδ | 1060 | 83 |
| 21-07-02 | 3.33 | 25-07-02 12.88h | 4.40 | 28 | 20-07-02 | 22:06 | No data | 1941 | H | S13E90 | X3.3/-- | 36 | βγδ | 880 | 26 |
| 14-08-02 | 2.83 | 15-08-02 16.88h | 1.59 | 26 | 14-08-02 | 2:30 | -0.20 | 1309 | 133 | N09W54 | M2.3/1N | 61 | βγ | 80 | 6 |
| 22-08-02 | 2.66 | 23-08-02 15.55h | 1.54 | 36 | 22-08-02 | 2:06 | -32.80 | 998 | H | S07W62 | M5.4/2B | 69 | βγδ | 1650 | 46 |
| 24-08-02 | 1.5 | 27-08-02 11.55h | 3.42 | 317 | 24-08-02 | 1:27 | 43.70 | 1913 | H | S02W81 | X3.1/1F | 69 | βγδ | 830 | 17 |
| 06-09-02 | 3.83 | 08-09-02 14.66h | 2.45 | 208 | 05-09-02 | 16:54 | 43.00 | 1748 | H | N09E28 | C5.2/SF | 102 | α | 20 | 1 |
| 09-11-02 | 15.42 | 11-11-02 5.77 h | 1.60 | 404 | 09-11-02 | 13:31 | 35.40 | 1838 | H | S12W29 | M4.6/2B | 180 | βγδ | 600 | 58 |
| 28-05-03 | 4.25 | 31-05-03 3.00h | 2.95 | 121 | 28-05-03 | 0:50 | 25.90 | 1366 | H | S07W20 | X3.6/-- | 365 | βγδ | 400 | 38 |
| 31-05-03 | 3 | 31-05-03 19.33h | 0.68 | 27 | 31-05-03 | 2:30 | -2.40 | 1835 | H | S07W65 | M9.3/2B | 365 | βγδ | 800 | 25 |
| 18-06-03 | 9.25 | 21-06-03 11.11h | 3.08 | 24 | 17-06-03 | 23:18 | -2.90 | 1813 | H | S07E55 | M6.8/-- | 386 | β | 190 | 13 |
| 26-10-03 | 18 | 28-10-03 11.75h | 1.74 | 466 | 26-10-03 | 17:54 | 4.80 | 1537 | >171 | N02W38 | X1.2/1N | 484 | βγδ | 1700 | 52 |
| 28-10-03 | 11.75 | 29-10-03 21.92h | 1.42 | 29500 | 28-10-03 | 11:30 | -105.20 | 2459 | H | S16E08 | X17./-- | 486 | βγδ | 2180 | 74 |
| 29-10-03 | 21.92 | 02-11-03 10.92h | 3.54 | 3300 | 29-10-03 | 20:54 | -146.50 | 2029 | H | S15W02 | X10./2B | 486 | βγδ | 2120 | 55 |
| 02-11-03 | 10.92 | 02-11-03 17.58h | 0.28 | 30 | 02-11-03 | 9:30 | -64.20 | 2036 | H | >SW90 | X9 | | No data | | |
| 02-11-03 | 17.58 | 04-11-03 22.25h | 2.19 | 1570 | 02-11-03 | 17:30 | -32.40 | 2598 | H | S14W56 | X8.3/2B | 486 | βγδ | 1900 | 99 |
| 04-11-03 | 22.25 | 07-11-03 22.66h | 3.02 | 353 | 04-11-03 | 19:54 | 434.80 | 2657 | H | S19W83 | X28./3B/ X18 | 486 | βγδ | 1430 | 16 |
| 21-11-03 | 16.5 | 23-11-03 0.02h | 1.31 | 14 | 20-11-03 | 8:06 | -44.70 | 669 | H | N01W08 | M9.6/2B | 501 | βγδ | 340 | 15 |
| 02-12-03 | 12.42 | 05-12-03 20.22h | 3.33 | 89 | 02-12-03 | 10:50 | 18.50 | 1393 | >150 | S11W90 | C8.0/-- | 508 | β | 140 | 8 |
| 11-04-04 | 6.17 | 12-04-04 23.11h | 1.71 | 35 | 11-04-04 | 4:30 | -77.60 | 1645 | 314 | S14W47 | C9.6/1F | 588 | β | 150 | 10 |
| 25-07-04 | 16.92 | 28-07-04 23.77h | 3.29 | 2090 | 25-07-04 | 14:54 | 7.00 | 1333 | H | N08W33 | M1.1/1F | 652 | βγδ | 1610 | 73 |
| 13-09-04 | 20 | 16-09-04 19.11h | 2.96 | 273 | 12-09-04 | 0:36 | 22.50 | 1328 | H | N03E49 | M4.8/2N | 672 | βγ | 260 | 49 |
| 19-09-04 | 18.17 | 21-09-04 11.11h | 1.71 | 57 | No data | | | | | N03W58 | M1.9/-- | 672 | β | 100 | 16 |
| 01-11-04 | 6.17 | 02-11-04 20.88h | 1.61 | 63 | 01-11-04 | 6:06 | -29.70 | 925 | 146 | >NW90 | c2.9 | 687 | β | 60 | 4 |
| 07-11-04 | 18.42 | 09-11-04 20.00h | 2.07 | 495 | 07-11-04 | 16:54 | -19.70 | 1759 | H | N09W17 | X2.0/-- | 696 | βγδ | 910 | 33 |
| 09-11-04 | 20 | 10-11-04 3.25h | 0.30 | 82 | 09-11-04 | 17:26 | -65.10 | 2000 | H | N08W51 | M8.9/2N | 696 | βγδ/ | 600 | 48 |
| 10-11-04 | 3.25 | 13-11-04 23.11h | 3.83 | 424 | 10-11-04 | 2:26 | -108.00 | 3387 | H | N09W49 | X2.5/3B | 696 | βγδ | 730 | 37 |
| 16-01-05 | 1.08 | 17-01-05 12.66h | 1.48 | 365 | 15-01-05 | 23:06 | -127.40 | 2861 | H | N15W05 | X2.6/-- | 720 | βδ | 1540 | 23 |
| 17-01-05 | 12.66 | 20-01-05 06.92h | 2.76 | 5040 | 17-01-05 | 9:54 | -159.10 | 2547 | H | N15W25 | X3.8/-- | 720 | βδ | 1630 | 37 |
| 20-01-05 | 6.92 | 23-01-05 22.44h | 3.64 | 1860 | 20-01-05 | 6:54 | 16.00 | 882 | H | N14W61 | X7.1/2B | 720 | βδ/ | 1220 | 31 |
| 13-05-05 | 18.75 | 16-05-05 22.44h | 3.15 | 3140 | 13-05-05 | 17:12 | No data | 1689 | H | N12E11 | M8.0/2B | 759 | β | 300 | 10 |
| 16-06-05 | 20.75 | 18-06-05 04.88h | 1.34 | 44 | No data | | | | | N08W90 | M4.0/SF | 775 | β | 230 | 4 |
| 14-07-05 | 13.08 | 17-07-05 14.25h | 3.05 | 134 | 14-07-05 | 10:54 | 198.00 | 2115 | H | N11W90 | X1.2/-- | 786 | βγδ | 370 | 8 |
| 17-07-05 | 14.25 | 19-07-05 15.33h | 2.05 | 22 | 17-07-05 | 11:30 | 59.20 | 1527 | H | >NW90 | | No data | | | |
| 27-07-05 | 22.42 | 02-08-05 13.11h | 5.61 | 41 | 27-07-05 | 4:54 | -75.40 | 1787 | H | N11E90 | M3.7/-- | 791 | β | 160 | 19 |
| 22-08-05 | 19.33 | 25-08-05 22.88h | 3.15 | 337 | 22-08-05 | 17:30 | 108.00 | 2378 | H | S13W65 | M5.6/1N | 798 | βγ | 560 | 18 |
| 07-09-05 | 20.17 | 13-09-05 23.00h | 6.12 | 1880 | No data | | | | | S11E77 | X17./3B | 808 | βγ | 10 | 1 |
| 13-09-05 | 23 | 16-09-05 22.66h | 2.99 | 183 | 13-09-05 | 20:00 | 11.50 | 1866 | H | S09E10 | X1.5/2B | 808 | βγδ | 840 | 52 |
| 06-12-06 | 23.25 | 13-12-06 02.92h | 6.15 | 1980 | 06-12-06 | 20:12 | No data | | H | S05E64 | X6.5/3B | 930 | β | 390 | 5 |
| 13-12-06 | 2.92 | 14-12-06 22.92h | 1.83 | 698 | 13-12-06 | 2:54 | -61.40 | 1774 | H | S06W23 | X3.4/4B | 930 | βγδ | 680 | 17 |
| 14-12-06 | 22.92 | 16-12-06 18.44h | 1.81 | 215 | 14-12-06 | 22:30 | -0.40 | 1042 | H | S06W46 | X1.5/-- | 930 | βγδ | 670 | 11 |
| 14-08-10 | 11.08 | 15-08-10 09.55h | 0.94 | 14 | 14-08-10 | 10:12 | -43.00 | 1205 | H | N17W52 | C4.4/-- | 1099 | β | 10 | 13 |
| 07-03-11 | 21.75 | 12-03-11 02h | 4.18 | 50 | 07-03-11 | 20:00 | -63.10 | 2125 | H | N31W53 | M3.7/-- | 1164 | βγδ | 640 | 25 |
| 21-03-11 | 4.17 | 22-03-11 19.11h | 1.62 | 14 | 21-03-11 | 2:24 | 8.00 | 1341 | H | >W90 | | 1169 | No Data | | |
| 07-06-11 | 7.33 | 09-06-11 16.66h | 2.39 | 72 | 07-06-11 | 6:49 | 0.30 | 1255 | H | S21W54 | M2.5/2N | 1226 | β | 80 | 8 |
| 04-08-11 | 4.5 | 07-08-11 12.22h | 3.32 | 96 | 04-08-11 | 4:12 | -41.10 | 1315 | H | N19W36 | M9.3/2B | 1261 | βγδ | 300 | 15 |
| 09-08-11 | 8.33 | 10-08-11 10.88h | 1.11 | 26 | 09-08-11 | 8:12 | -40.60 | 1610 | H | N17W69 | X6.9/2B | 1263 | βγδ | 450 | 13 |
| 22-09-11 | 17.92 | 29-09-11 14.66h | 6.86 | 35 | 22-09-11 | 10:48 | -68.30 | 1905 | H | N09E89 | X1.4/-- | 1302 | βγ | 480 | 2 |
| 26-11-11 | 8.25 | 29-11-11 8.88h | 3.03 | 80 | 26-11-11 | 7:12 | 9.00 | 933 | H | N17W49 | C1.2/-- | 1353 | αγ | 10 | 1 |
| 23-01-12 | 4.75 | 27-01-12 18.92h | 4.59 | 6310 | 23-01-12 | 4:00 | 28.00 | 2175 | H | N28W21 | M8.7/-- | 1402 | βγ | 370 | 8 |
| 27-01-12 | 18.92 | 01-02-12 8.88h | 4.58 | 795 | 27-01-12 | 18:27 | 165.90 | 2508 | H | N27W71 | X1.7/1F | 1402 | βγ | 270 | 9 |
| 07-03-12 | 2.83 | 13-03-12 18.08h | 6.63 | 6530 | 07-03-12 | 0:24 | -88.20 | 2684 | H | N17E27 | X5.4/-- | 1429 | βγδ | 1120 | 25 |
| 13-03-12 | 18.08 | 16-03-12 16.00h | 2.91 | 469 | 13-03-12 | 17:36 | 45.60 | 1884 | H | N17W66 | M7.9/-- | 1429 | βγδ | 380 | 14 |
| 17-05-12 | 1.92 | 19-05-12 20.44h | 2.77 | 255 | 17-05-12 | 1:48 | -51.80 | 1582 | H | N11W76 | M5.1/1F | 1476 | βγδ | 230 | 3 |
| 26-05-12 | 23.42 | 28-05-12 14.22h | 1.62 | 14 | 26-05-12 | 20:57 | -159.20 | 1966 | H | >W90 | | No data | | | |
| 14-06-12 | 23.42 | 17-06-12 12.88h | 2.56 | 15 | 14-06-12 | 14:12 | -1.20 | 987 | H | S17E06 | M1.9/1N | 1504 | βγδ | 560 | 25 |
| 07-07-12 | 0.08 | 08-07-12 18.17h | 1.75 | 25 | 06-07-12 | 23:24 | -56.10 | 1828 | H | S13W59 | X1.1/-- | 1515 | βγδ | 670 | 56 |
| 08-07-12 | 18.17 | 10-07-12 16.22h | 1.91 | 19 | 08-07-12 | 16:54 | -117.20 | 1497 | 157 | S17W74 | M6.9/1N | 1515 | βγ | 780 | 43 |
| 12-07-12 | 17.42 | 15-07-12 10.22h | 2.70 | 96 | 12-07-12 | 16:48 | 195.60 | 885 | H | S15W01 | X1.4/-- | 1520 | βγδ | 1320 | 48 |
| 17-07-12 | 15.5 | 19-07-12 6.66h | 1.63 | 136 | 17-07-12 | 13:48 | 63.50 | 958 | 176 | S28W65 | C9.9/1F | 1520 | βγδ | 430 | 12 |
| 19-07-12 | 6.66 | 23-07-12 8.00h | 4.05 | 80 | 19-07-12 | 5:24 | -8.00 | 1631 | H | S13W88 | M7.7/-- | 1520 | βγδ | 300 | 5 |
| 23-07-12 | 8 | 26-07-12 19.33h | 3.47 | 12 | 23-07-12 | 2:36 | -24.60 | 2003 | H | >W90 | | No data | | | |
| 01-09-12 | 1.42 | 05-09-12 7.11h | 4.24 | 60 | 31-08-12 | 20:00 | 2.00 | 1442 | H | S25E59 | C8.1/2F | 1562 | α | 10 | 1 |
| 28-09-12 | 1.33 | 30-09-12 20.00h | 2.78 | 28 | 28-09-12 | 0:12 | -27.10 | 947 | H | N06W34 | C3.7/1F | 1577 | βγ | 15 | 3 |
| 15-03-13 | 19.66 | 18-03-13 4.22h | 2.36 | 16 | 15-03-13 | 7:12 | 25.80 | 1063 | H | N11E12 | M1.1/1F | 1692 | α | 200 | 2 |
| 11-04-13 | 8.42 | 14-04-13 20.00h | 3.48 | 114 | 11-04-13 | 7:24 | -8.10 | 861 | H | N09E12 | M6.5/3B | 1719 | βγ | 190 | 17 |
| 15-05-13 | 6.58 | 21-05-13 15.33h | 6.36 | 42 | 15-05-13 | 1:48 | -52.10 | 1366 | H | N12E64 | X1.2/2N | | βγδ | 310 | 5 |
| 22-05-13 | 14.33 | 26-05-13 3.11h | 3.53 | 1660 | 22-05-13 | 13:25 | -13.20 | 1466 | H | N15W70 | M5.0/-- | 1745 | βγδ | 120 | 5 |
| 23-06-13 | 8.5 | 25-06-13 9.11h | 2.03 | 14 | 21-06-13 | 3:12 | 1.50 | 1900 | >207 | S16E73 | M2.9/-- | 1778 | αγ | 110 | 1 |
| 30-09-13 | 0.42 | 03-10-13 15.11h | 3.61 | 182 | 29-09-13 | 22:12 | -5.30 | 1179 | H | N17W29 | C1.6/-- | 1850 | βγ | 100 | 7 |
| 28-12-13 | 19 | 30-12-13 17.33h | 1.93 | 29 | 28-12-13 | 17:36 | -26.70 | 1118 | H | >W90 | C9 | 1936 | | | |
| 06-01-14 | 8.25 | 07-01-14 19.92h | 1.49 | 42 | 06-01-14 | 8:00 | -7.10 | 1402 | H | >W90 | C2.1/-- | 1936/1937 | βγ | 160 | 21 |
| 07-01-14 | 19.92 | 14-01-14 2.00h | 6.25 | 1026 | 07-01-14 | 18:24 | -60.80 | 1830 | H | S15W11 | X1.2/-- | 1944 | βγδ | 1415 | 11 |
| 20-02-14 | 8.25 | 21-02-14 3.55h | 0.80 | 22 | 20-02-14 | 8:00 | -9.50 | 948 | H | S15W73 | M3.0/-- | 1976 | α | 180 | 5 |
| 25-02-14 | 3.83 | 05-03-14 20.00h | 8.67 | 24 | 25-02-14 | 1:25 | -158.10 | 2147 | H | S12E82 | X4.9/-- | 1990 | α | 250 | 2 |
| 18-04-14 | 13.66 | 21-04-14 10.00h | 2.85 | 58 | 18-04-14 | 13:25 | 13.50 | 1203 | H | S20W34 | M7.3/-- | 2036 | βγ | 510 | 34 |
| 10-09-14 | 21.58 | 14-09-14 4.44h | 3.29 | 126 | 10-09-14 | 18:00 | -51.60 | 1267 | H | N14E02 | X1.6/-- | 2158 | βγδ | 420 | 25 |
| 01-11-14 | 13.92 | 04-11-14 6.44h | 2.69 | 11 | 01-11-14 | 6:00 | -1.00 | 740 | >160 | N27W79 | | No data | | | |
| 18-06-15 | 4.58 | 20-06-15 16.66h | 2.50 | 17 | 18-06-15 | 1:25 | 27.70 | 1714 | 195 | S16W81 | M1.2/-- | 2365 | α | 60 | 1 |
| 21-06-15 | 4.08 | 25-06-15 10.08h | 4.25 | 1066 | 21-06-15 | 2:36 | 21.20 | 1366 | H | N12E16 | M2.6/-- | No data | | | |
| 25-06-15 | 10.08 | 30-06-15 8.22h | 4.92 | 22 | 25-06-15 | 8:36 | -24.80 | 1627 | H | N09W42 | M7.9/-- | 2371 | βγ | 740 | 29 |
| 29-10-15 | 3.08 | 30-10-15 20.66h | 1.73 | 24 | 29-10-15 | 2:36 | -7.50 | 530 | 202 | >W90b | | 2434 | No data | | |
| 02-01-16 | 0.25 | 03-01-16 08.00h | 1.32 | 22 | 01-01-16 | 23:24 | 12.70 | 1730 | H | S25W82 | M2.3/-- | 2473 | β | 130 | 8 |
| 14-07-17 | 4.66 | 16-07-17 13.77h | 2.38 | 22 | 14-07-17 | 1:25 | -0.10 | 1200 | H | S06W29 | M2.4/-- | 2665 | β | 440 | 26 |
| 04-09-17 | 22.5 | 06-09-17 12.58h | 1.59 | 210 | 04-09-17 | 20:12 | 47.50 | 1418 | H | S10W12 | M5.5/-- | 2673 | βγ | 130 | 12 |
| 06-09-17 | 12.58 | 09-09-17 19.11h | 3.27 | 844 | 06-09-17 | 12:24 | -0.30 | 1571 | H | S08W33 | X9.3/-- | 2673 | βγδ | 880 | 33 |
| 10-09-17 | 16.42 | 15-09-17 5.11h | 4.53 | 1490 | 10-09-17 | 16:00 | -232.00 | 3163 | H | >W90b | X8.2/-- | 2673 | βγδ | 530 | 8 |